# How Nanobubbles Nucleate at a Hydrophobic/Water Interface

Ing-Shouh Hwang*, Chung-Kai Fang, Hsien-Chen Ko, Chih-Wen Yang, and Yi-Hsien Lu

Institute of Physics, Academia Sinica, Nankang, Taipei, Taiwan

*ishwang@phys.sinica.edu.tw

**ABSTRACT** Experimental investigations of hydrophobic/water interfaces often return controversial results, possibly due to the unknown role of gas accumulation at the interfaces. Here, during advanced atomic force microscopy of the initial evolution of gas-containing structures at a highly ordered pyrolytic graphite/water interface, a fluid phase first appeared as a circular wetting layer ~0.3 nm in thickness and was later transformed into a cap-shaped nanostructure (an interfacial nanobubble). Two-dimensional ordered domains were nucleated and grew over time outside or at the perimeter of the fluid regions, eventually confining growth of the fluid regions to the vertical direction. We determined that interfacial nanobubbles and fluid layers have very similar mechanical properties, suggesting low interfacial tension with water and a liquid-like nature, explaining their high stability and their roles in boundary slip and bubble nucleation. These ordered domains may be the interfacial hydrophilic gas hydrates and/or the long-sought chemical surface heterogeneities responsible for contact line pinning and contact angle hysteresis. The gradual nucleation and growth of hydrophilic ordered domains renders the original homogeneous hydrophobic/water interface more heterogeneous over time, which would have great consequence for interfacial properties that affect diverse phenomena, including interactions in water, chemical reactions, and the self-assembly and function of biological molecules.

Gases dissolved in water tend to accumulate at the interfaces between hydrophobic solids and water to form cap-shaped structures that are nanometers in height; these structures are known as interfacial nanobubbles (INBs) or surface nanobubbles[1-9]. INBs have attracted much attention because of their potential implications for various interfacial phenomena and technical applications, such as long-range attractive forces between hydrophobic surfaces in solutions[10], liquid slippage at hydrophobic walls[11-13], the stability of colloidal systems[14], and bio-molecular adsorption[15]. INBs are also proposed to be the gas micronuclei that are responsible for bubble formation at solid/water interfaces[16]. However, the nature of INBs remains unknown, and the mechanisms responsible for the above phenomena are not clear.

To date, most studies have focused on why INBs exhibit high stability and why they adopt a rather flat morphology. The lifetime of an INB should be much less than 1 ms based on classical diffusion theory[17,18], but experimental observations have indicated that INBs can persist for days, which is at least 10 orders of magnitude longer than the theoretical prediction[5,6]. Although several models have been proposed to explain this unexpected stability, no consensus has been reached. A very recent model was based on the pinning of the three-phase INB-water-surface contact line[19,20], which was attributed to omnipresent chemical and geometrical surface heterogeneities[20] of unknown origin. Such surface heterogeneities were also suggested to lead to contact angle hysteresis, the difference between an advancing and a receding contact angle, for water droplets on solid surfaces[21-23].

Another fundamental but rarely addressed issue is the mechanism by which INBs nucleate at hydrophobic/water interfaces. Here, we used advanced atomic force microscopy (AFM) to investigate the initial formation of gas-containing structures at the interface between water and a mildly hydrophobic solid, highly ordered pyrolytic graphite (HOPG). Our observations, which were conducted in the frequency-modulation (FM) and PeakForce (PF) modes, provide important insights into the behaviors of dissolved gases in water and at hydrophobic/water interfaces, including formation of INBs, their nature, their high stability, the pinning of the three-phase contact lines, boundary slip, and bubble nucleation. This new understanding highlights new directions for further quantitative investigation of these behaviors under various conditions and at various interfaces, which will allow accurate prediction and control of these behaviors. These advances will strongly impact research and technology in diverse fields, such as physics, chemistry, biology, and medicine.

## RESULTS AND DISCUSSION

**Formation of gas-containing structures at a HOPG/water interface.** Figure 1 shows an example of our observations after chilled water (see Methods) was deposited onto a freshly cleaved HOPG surface. Several regions, several of which are numbered in Figure 1a, exhibited a thin circular layer covering the HOPG substrate. Notice that several circular layers had a diameter as large as ~1 μm, and some had edges bounded by defects such as substrate step edges. The height of the circular layer was 0.3-0.4 nm (Figure S1a). In addition, bright speckles were scattered at the interface outside the circular layered regions (Figure 1a). Higher-resolution images showed that the speckles were small domains with an ordered row-like structure (Figures 1b and c) that increased in coverage of the interface over time (Figure 1d-h). This structure and its nucleation and growth behaviours resemble the ordered pattern reported in our earlier studies of the HOPG/water interface using pre-degassed water under air or nitrogen[24-27].

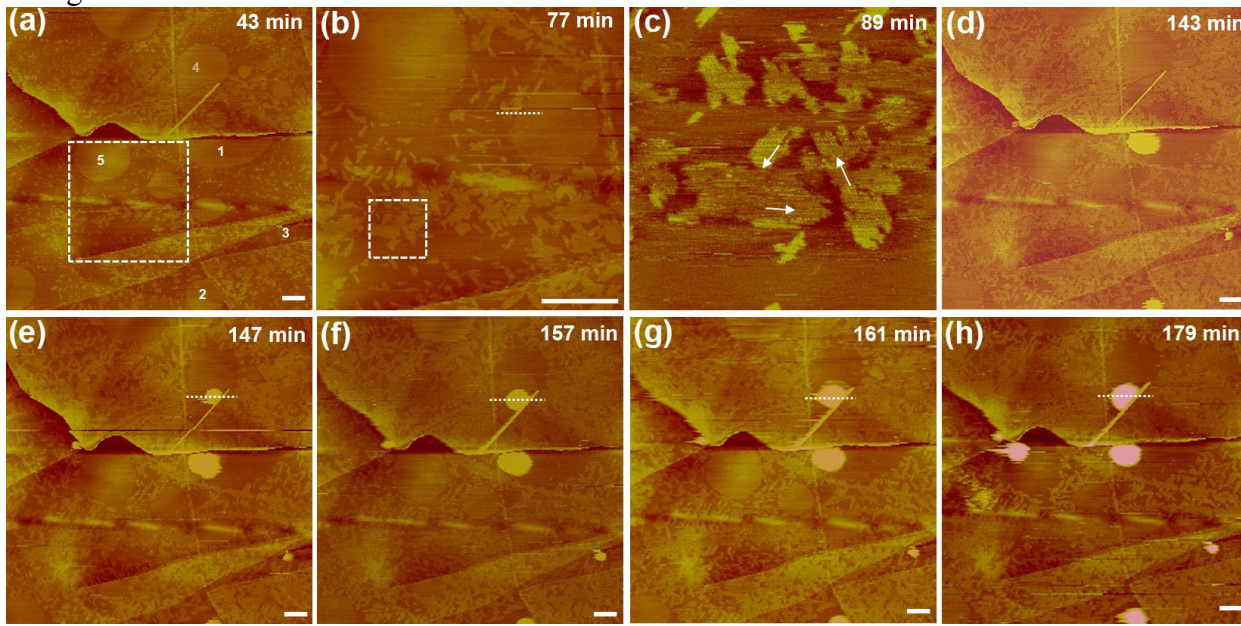

**Figure 1. Topographic images of the formation of gas-containing structures at a HOPG/water interface acquired with FM-AFM. Time points are relative to the time of water deposition (t=0). Scale bar, 400 nm. Numbered regions in (a) exhibit a circular, thin layer covering the HOPG substrate that changed in morphology over time. The numbers in panel (a) are not shown in the other panels. (b) A higher-resolution image acquired inside the dashed box in (a) at a later time. The height profile measured along a white dashed line is shown in Figure S1a. (c) A high-resolution image of bright speckles (inside the dashed box in (b) at a later time), consisting of domains of a row-like pattern with three orientations (arrows) that**

**are parallel to the orientations of the HOPG substrate. The domains with approximately horizontal row orientation (the slow scan direction) cannot be clearly resolved. A cap-shaped structure appeared in Region 4 in (e) and exhibited growth in both the vertical and lateral dimensions (f)-(h). The height profiles along a white dashed line across the INB from (e) to (h) are depicted in Figure S1b.**

At t=143 min, the thin layers in regions 1-3 transformed into a cap-shaped INB (Figure 1d). Four minutes later, a small cap-shaped protrusion appeared on the thin layer in region 4 (Figure 1e). The thin layer gradually receded, revealing its fluid nature, while the protrusion grew in both the lateral and vertical dimensions (Figure 1e-h and Figure S1b). The ordered domains covered more of the interface surface over time and confined the lateral growth of fluid layers at late stages. After the edges of the fluid layer receded, a region of bare HOPG surface bounded by ordered domains and substrate step edges was exposed (Figure 1e-h), indicating that no ordered structure was formed inside the fluid layer. In the area near the fluid layer in region 4, several smaller fluid layers displayed a circular shape at t=43 min but grew laterally until their boundaries were confined by the ordered domains (Figure S2).

Another AFM observation revealed two thin circular layers >1 μm in diameter; both displayed a very rapid (<1 s) transition from a fluid layer to a cap-shaped INB (Figure S3). Similar to the observation in Figure 1, a large, bare HOPG surface was exposed directly after the transition, but small ordered domains gradually nucleated inside the bare HOPG regions (Figure S3e), and the two INBs grew in the vertical and lateral directions (Figure S3g,h). We note that the ordered domains exhibited a smaller dissipation than the bare HOPG surface (Figures S2 and S3), consistent with our previous observations[24-28].

We detected a different transformation process for smaller fluid regions when chilled water was rapidly heated to 45 °C before deposition (Figure 2). At t=20 min and t=23 min, gas adsorption at the interface was evident in the stiffness and adsorption maps, but was barely detectable in the topographic images acquired in PF mode (white arrow, Figure 2a and 2b). This difference occurred because the AFM tip usually penetrates into fluid structures to a certain depth that depends on the hydrophobicity of the tip[29], in order to achieve a positive pre-set peak force due to the capillary force between the tip and the fluid structure[28]. When the fluid structure is too thin, the tip traces the profile of the underlying stiff structure (the substrate, in this case). At t=26 min, the topographic image began to reveal the presence of a circular layer 0.4-0.5 nm in thickness (white arrow, Figure 2c). At t=28 min, the apparent height of the fluid layer continued to grow, and a small protrusion appeared near the centre of the layer (Figure 2d). A cap-shaped INB eventually formed (Figure 2f-l).

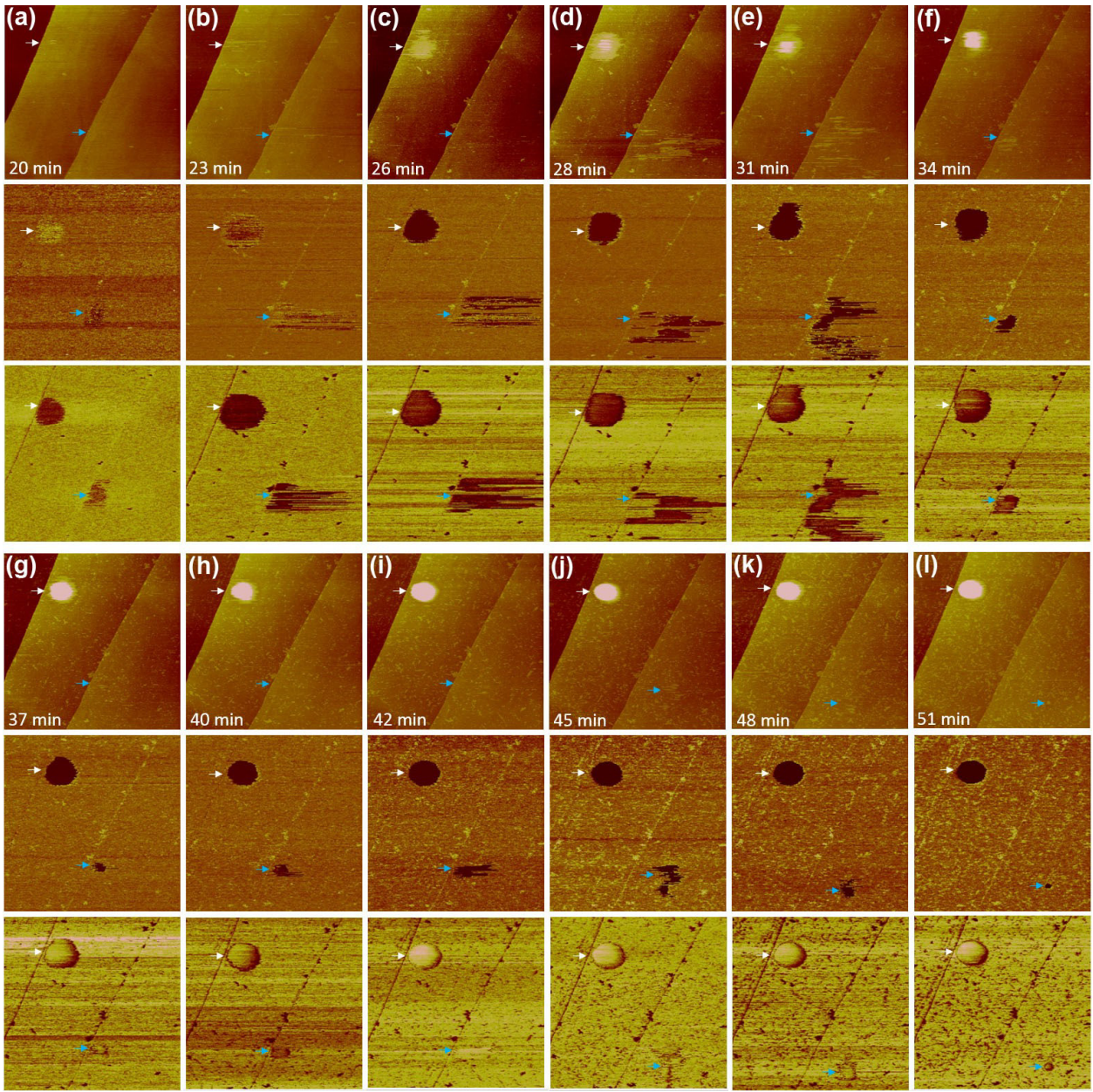

**Figure 2.** AFM images (PF mode, 250 pN) of the transition from a 2D fluid layer to a 3D structure. The scan area is 2 μm × 2 μm. Chilled water was rapidly heated to 45 °C and deposited on a freshly cleaved HOPG substrate. Images were acquired continuously with a scanning rate of ~3 min per image. The top, middle, and bottom rows of each panel depict the topographic, stiffness (Young's modulus), and adhesion maps, respectively. Two fluid regions are highlighted with white and blue arrows.

The stiffness maps (middle rows in Figure 2) indicate that the fluid phase is softer than the other regions of the interface; thus, the stiffness maps enable easy detection of the fluid regions. However, the quantitative stiffness value may be not accurately extracted for a structure <2 nm in thickness, a limitation that occasionally led to a reversal of the stiffness contrast of the fluid structures (Figure 2a). Adhesion contrast between the fluid regions and bare HOPG regions varies from tip to tip, but still allows distinction between these two regions. The stiffness and adhesion maps did not provide enough contrast to distinguish between the fluid layer and the cap-shaped structure (Figures 2d,e), suggesting that these two structures share a very similar nature. The ordered domains are visible in Figure 2, but they appear as white and dark speckles in the topographic and adhesion maps, respectively, due to the large scan area of the images. The ordered domains consistently exhibited a higher stiffness than the fluid structures (Figure 2). However, they appeared slightly stiffer or softer than the bare HOPG regions, depending on the tip and operating conditions, again because the ordered domains were <2 nm in thickness and their stiffness values could not be accurately extracted.

The apparent height of the fluid region highlighted with a white arrow in Figure 2 gradually increased over time (Figure 3a). The real thickness of the fluid structures (the thin layer and the cap-shaped structure) should be larger than the apparent height due to the effect of tip-penetration depth; thus, the fluid layer in Figures 2c and 2d is more than one molecule thick. After a protrusion appeared at t=28 min, both the lateral size and the apparent height of the protrusion continued to grow until t=37 min, after which growth mainly occurred in the vertical dimension (the lateral size remained roughly the same) (Figure 3b). The growth behavior of INBs after t=37 min resembled that reported by Zhang et al. (who changed the air-supersaturation level), which Zhang et al. attributed to the pinning of the three-phase contact line[30]. The apparent height of the fluid region indicated with a white arrow in Figure 2 continued to increase over time (Figure S4); the lateral area initially increased, but gradually decreased at later times (Figure S4), in contrast to the rapid reduction in the lateral size of the fluid layer evident in Figure 1 and Figure S3.

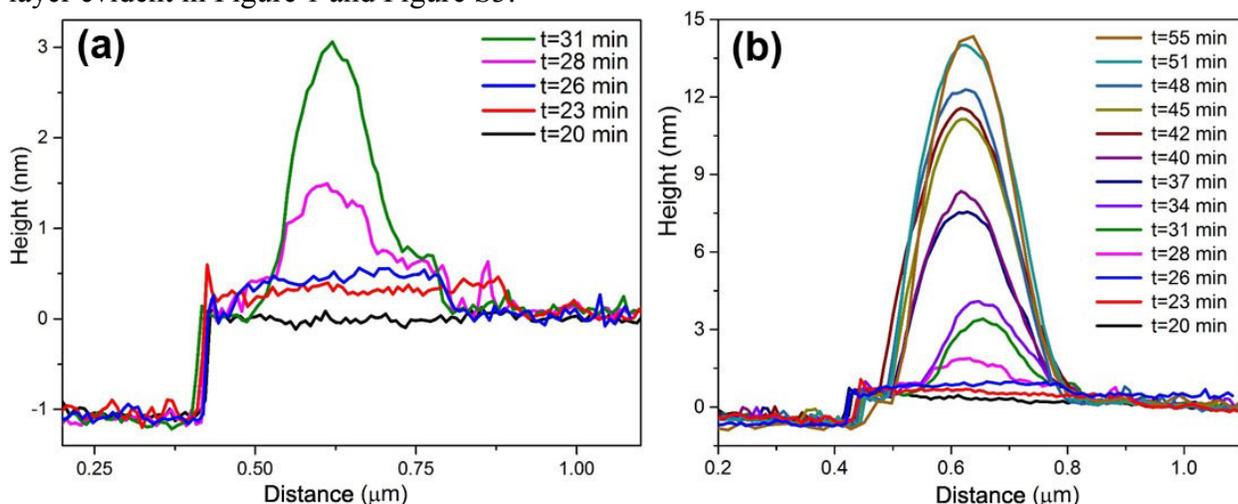

**Figure 3 Apparent height profiles of the fluid structure indicated by a white arrow in Figure 2. (a) Profiles acquired from t=20 min to t=31 min, during the initial stages of INB formation. (b) Overall profiles acquired from t=20 min to t=55 min.**

Another region of gas adsorption is visible in the adhesion and stiffness maps in Figure 2 (blue arrow). The lateral shape of this region was irregular and changed from frame to frame (Figure 2), suggesting that the gas structure was highly mobile in this region during tip scanning. This mobility appeared to slow at t=48 min (Figure 2k), and the structure transformed into a small INB at t=51 min (Figure 2l). The high mobility of this gas structure seems to be related to its small size; this gas structure is barely detectable in the topographic images in Figure 2.

Enlarging the scan area of Figure 2 revealed additional INBs (Figure S5a), indicating that the 2D-to-3D transition of fluid regions is spontaneous. However, we cannot rule out the possibility that the scanning tip may facilitate this transition. Ordered domains of the row-like structure were evident in regions outside the INBs (Figure S5b).

We identified an interesting case of relatively fast nucleation and growth of the ordered domains (Figure 4). Fluid layers (some indicated with colored arrows in Figure 4a) and low-coverage ordered domains appeared in the stiffness and adhesion maps at t=8 min (Figure 4a). Confinement of the fluid regions occurred at t=12 min (Figure 4b). The fluid regions seemed to be present at a lower height than the surrounding ordered domains and were barely distinguishable from the bare HOPG regions in the topographic images (Figure 4a,b). Portions of the outer parts of the fluid regions transformed into ordered domains, decreasing the lateral size of the fluid regions (Figure 4b-d). Cap-shaped protrusions were evident in the topographic images at t~50 min (data not shown). At t=352 min, all regions highlighted by colored arrows in Figure 4 transformed into cap-shaped structures (Figure 4e), with the exception of the region denoted by a green arrow in Figure 4a.

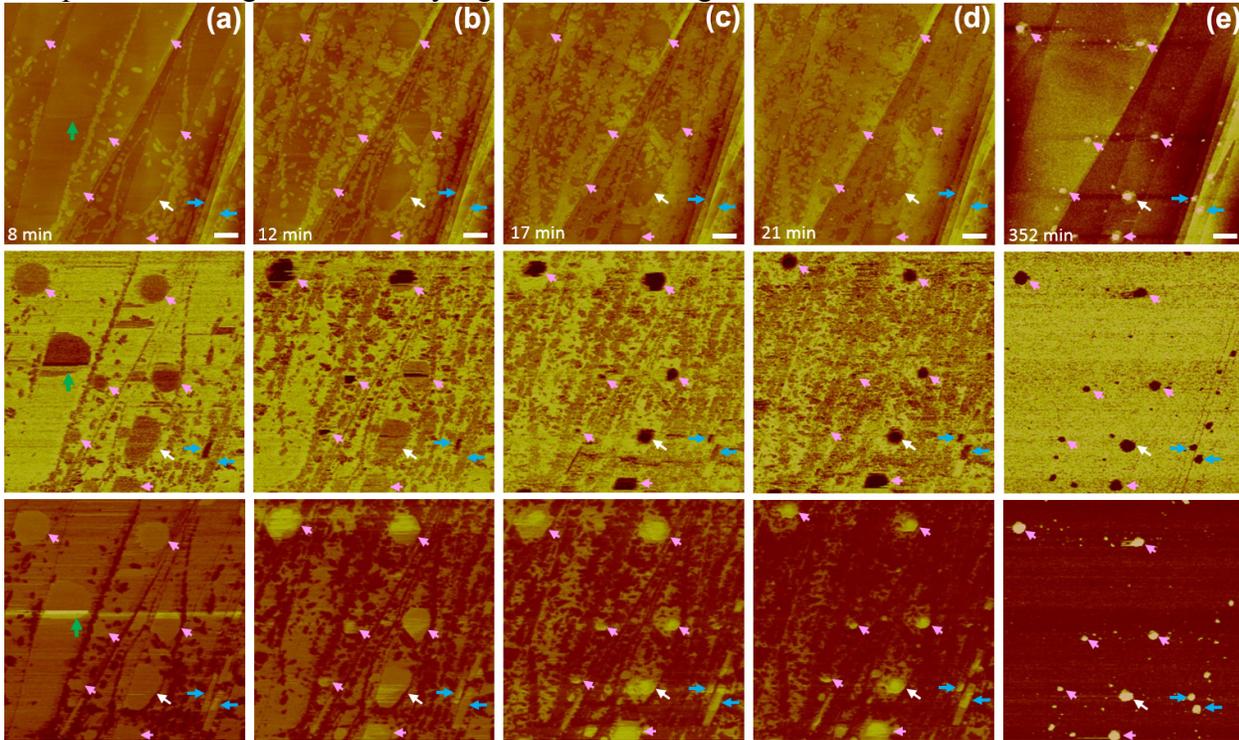

**Figure 4 PF (220 pN) images of gas-containing structures at a HOPG/water interface after the deposition of chilled water. Top, middle, and bottom rows contain height, stiffness, and adhesion maps, respectively. Scale bar, 500 nm. The green arrow in (a) highlights a fluid layer that disappeared at a later time (b). The white, blue, and pink arrows indicate fluid regions that later transformed into a cap-shaped structure (e). Two blue arrows indicate fluid regions that were confined by two step edges, yielding rectangular gas structures (a-d). Note that the ordered domains exhibited less adhesion than the bare HOPG surface and the fluid structures. In (e), the entire interface (except for the INBs) was dark on the adhesion maps (bottom row) and bright on the stiffness maps (middle row), indicating that the interface was nearly completely covered by the ordered domains. The regions of the fluid structures can be distinguished easily in the stiffness maps because they exhibited less stiffness than the ordered domains and HOPG substrate. This AFM tip was relatively hydrophobic, yielding a penetration depth for the fluid regions that was larger than that obtained with the tips used to acquire the data in the other figures presented in this work. A cap-shaped structure is visible in a topographic image only when its height is larger than the tip penetration depth. In addition, it is difficult to distinguish between INBs and fluid layers from the stiffness and adhesion maps. Thus, the INBs may have formed well before we detected the cap-shaped structures in the topographic (height) maps.**

We acquired a higher-resolution topographic image around the cap-shaped structure indicated with a white arrow in Figure 4 at t=475 min (Figure 5a). At this time point, the INB was already surrounded by several layers of ordered structures. When we scanned the INB at a higher peak force, the INB appeared as a depression surrounded by multi-layer ordered domains (Figure 5b). We previously demonstrated that a higher peak force results in a lower surface profile of an INB because the tip penetrates more deeply into the fluid structure[28]. Figure 5b shows that only bare HOPG substrate, with no ordered structures, was present under the INB. The edges of the surrounding ordered structures were faceted and roughly hexagonal (Figure 5b). The thickness of the ordered structures was larger at the perimeter of the INB than at other locations away from the INB, suggesting that the formation of ordered structures is more favorable next to a fluid structure. Images acquired at later times at this high peak force revealed the growth of the ordered structures at the outer boundaries of the INB (Figure 5c) and clearly visualized strong confinement of the INB by the surrounding ordered structures.

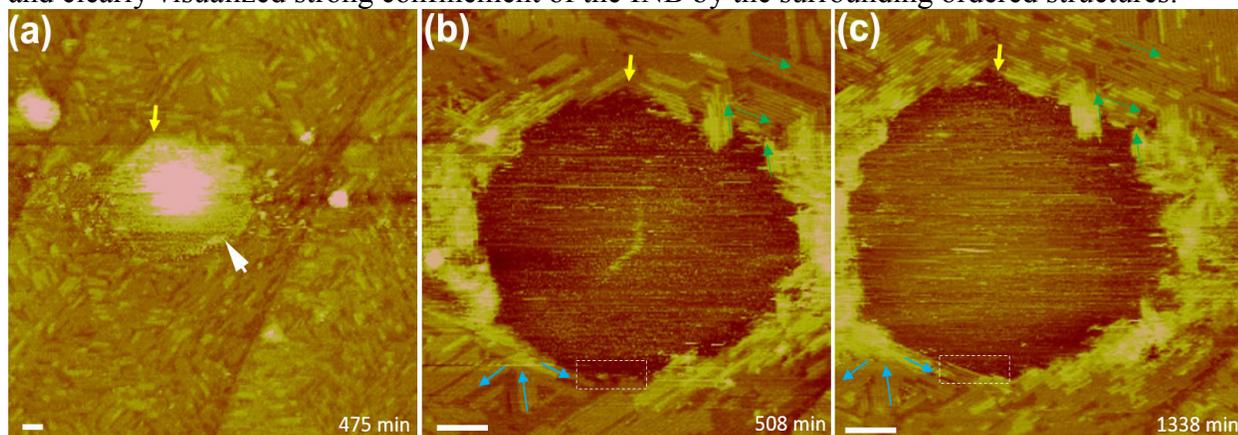

**Figure 5** High-resolution topographic images (PF mode) around the region indicated with a white arrow in Figure 4 acquired at later times. The peak force was set to 220 pN for (a) and 660 pN for (b) and (c). The yellow arrow serves as a marker for comparison among the three panels. Scale bar, 70 nm. The images in (b) and (c) were acquired with the slow scan direction rotated 20° clockwise relative to that in (a); the row-like patterns could not be well resolved when the row orientation was parallel with the slow scan direction. Blue and green arrows are indicated in (b) and (c) for comparison of structural changes over time. Growth of more layers of ordered structures at the outer edges of the INB is evident in (c). The region outlined by a white dashed box in (b) and (c) highlights the growth of the ordered structures at a boundary of the INB.

**Nature of the fluid layers and INBs at the HOPG/water interface.** The surprising formation of a large wetting layer (1 μm in diameter) one molecule in thickness (Figure 1 and Figure S3) indicates the presence of an adsorbed gas film that is in a liquid-like state rather than in a vapor phase. Most importantly, this observation revealed that there is a tendency for the adsorbed gas structure to spread as much as possible along the HOPG/water interface (Supporting Information Additional Discussion 1). This excellent wetting and spreading behavior indicates that the interfacial tension (interface energy per unit area) between the adsorbed gas film and water, $\sigma_{\text{film-water}}$, should be considerably smaller than the surface tension of water ($\sigma_{\text{vapor-water}}$=72 mN/m) (Supporting Information Additional Discussion 2). The high surface tension of water is the driving force that retains semispherical gas bubbles microns in size or larger at solid/water interfaces.

Based on several observations, we conclude that INBs are liquid-like gas condensates at the solid/water interface, rather than surface bubbles with low-density gases in their vapor phase. First, we observed a relatively continuous height increase during the 2D-to-3D transition of the fluid structure (Figures 2 and 3), without an abrupt and large volume change corresponding to a liquid-vapor phase transition. Second, our PF measurements (Figures 2 and 4) indicate that the mechanical properties (stiffness and adhesion) of the fluid layers and the cap-shaped structures are nearly indistinguishable. Third, we determined that a cap-shaped structure can form on top of a fluid layer and persist for more than 10 min (Figures 1 and 2), demonstrating that these two structures have similar chemical potentials and a similar nature. Since INBs exhibit good wetting with the interface with a very small contact angle (5°~25° measured from the gas side[5-8]), we expect that the interfacial tension between an INB and water is significantly smaller than the surface tension of water.

The low interfacial tension of INBs may account for the surprisingly high stability of INBs observed experimentally. It also solves two great mysteries in the science of solid/water interfaces: boundary slip and the nucleation of gas bubbles in water. An INB may serve as a highly deformable liquid-like layer of low viscosity that reduces the drag of flowing water over a hydrophobic wall; this function would explain the mechanism of the boundary-slip effect proposed by Vinogradova[31], who also proposed that gas bubbles with high surface tension would not deform easily and thus could not contribute to drag reduction[32]. Consistent with this hypothesis, gas bubbles trapped at solid/water interfaces acted as anti-lubricants and increased drag[33]. INBs were suggested to be responsible for boundary slip on hydrophobic surfaces, but they were generally considered to be air bubbles on the nanometer scale[12,13]. Our demonstration of the drastically different natures of INBs and gas bubbles resolves this contradiction. The low interfacial tension of INBs may also underlie the nucleation of gas bubbles in water, and thus large INBs may be the long-sought micronuclei for gas bubble formation in water (Supporting Information Additional Discussion 3). Quantitative measurement of the interfacial tension between an INB and water, which was beyond the scope of the present investigation, will enable further interrogation of these hypotheses.

From the thermodynamic point of view, the high surface tension of water makes small gas bubbles energetically unfavorable, because the interfacial energy contributes to a larger portion of the total energy (bulk energy plus interfacial energy) as the size of a gas bubble decreases. Gas bubbles microns in size or larger exhibit such a trend, but INBs do not—INBs are extremely stable. We suspect that the high surface tension of water may be the thermodynamic driving force that condenses very small gas bubbles into gas condensates, which become energetically favorable due to a much lower interfacial tension with water and a smaller interfacial area[25,26,28]. As described above, the low interfacial tension of INBs with water solves several outstanding mysteries, including boundary slip and the nucleation of gas bubbles. The low interfacial tension is also consistent with several recent experimental observations. First, transmission electron microscopy of merging INBs[34] showed that INBs are highly deformable and can exhibit a shape that strongly deviates from a semisphere, e.g. a dumbbell shape, for tens of seconds, which is considerably longer the merging process for surface bubbles with a diameter of tens of microns to centimeters (< 1 s). Second, INBs did not burst when a macroscopic surface bubble (containing water vapor) grew and moved across the INBs[35]; instead, a water film wet the INBs and nucleated a growing water droplet through vapor condensation. This process reflects the favorable affinity of INBs for water, which is consistent with the low interfacial tension proposed here but not with the high interfacial tension associated with a gas bubble in water. Third, giant INBs with a lateral diameter of 8-10 μm were reported in recent years[36,37]. They exhibited a rather flat and stable morphology similar to that of smaller INBs[36]. Surface gas bubbles with this lateral diameter are visible on typical optical microscopy and exhibit a semispherical shape with a much larger height/diameter ratio than INBs. In addition, these small gas bubbles are unstable and grow or shrink within a few seconds, depending on the dissolved gas concentration. In contrast, the morphology of INBs changes only slightly many hours after the change in gas concentration[30], indicating very inefficient gas transfer between INBs and the surrounding water. The observations of giant INBs indicate that it is their nature, rather than their size, that differentiates INBs from typical surface gas bubbles. Thus, we suggest that "nanobubble" is not an appropriate name for these cap-shaped fluid structures that consist of liquid-like gas condensates at solid/water interfaces.

**Interfacial gas hydrate phases**. We suspect that the fluid structures observed here are aggregates of pure gas molecules (nitrogen and oxygen) confined at the HOPG/water interface, where there are only weak van der Waals interactions among these gas molecules. Our observations indicate that ordered domains appear only in the areas outside or at the perimeter of the fluid regions (Figures 1, 2, 4, and 5 as well as Figures S2, S3, and S5), suggesting that the presence of water is essential for formation of these ordered structures. The rather slow nucleation and growth processes of the ordered domains relative to the appearance of the fluid layers in water (Figures 1, 2, and 4 as well as Figures S2 and S3) suggests that the formation of ordered domains requires a complicated bonding arrangement of water and gas molecules, rather than a simple aggregation of pure gas molecules at the interface. Thus, the gas-containing ordered structures detected here bear some resemblance to bulk gas hydrates in which water molecules are linked through hydrogen bonding and create cavities that harbor gas molecules[38,39].

Hydrogen-bonded networks of water can stabilize gas-containing ordered structures at the interface, explaining why the ordered structures exhibit mechanical properties that are very different from those of the fluid structures (Figures 2 and 4 as well as Figures S2 and S3). Interfacial gas-hydrate phases would be more hydrophilic than bare HOPG surface, resulting in less adhesion and dissipation on AFM (Figures 2 and 4 as well as Figures S2 and S3). Hydrophilic gas-hydrate domains act as barriers to the lateral growth of fluid structures, which preferentially adsorb onto bare HOPG regions (hydrophobic regions) in water. Similarly, the step edges of the HOPG substrate confined the lateral growth of the fluid layers, probably because they are hydrophilic due to broken bonds. These hydrophilic structures may serve as pinning sites for the lateral confinement of the fluid layers and INBs. Pinning of the three-phase nanobubble-water-surface contact line was previously attributed to omnipresent chemical and geometrical surface heterogeneities[20]. Such surface heterogeneities were also considered to underlie contact angle hysteresis[21-23]. Our observations of substrate step edges and the formation of ordered domains provide microscopic evidence of the origin of such surface heterogeneities. This evidence is remarkable, because we began our experiments with a relatively homogeneous and clean HOPG surface in pure water (see Methods). Gases dissolved in water segregate to the interface, rendering the interface highly heterogeneous on the nanometer scale. Therefore, future experimental studies (with probe size >100 nm) and theoretical modelling of the interface will need to consider the heterogeneous nature of the hydrophobic/water interface demonstrated here.

We have demonstrated that the ordered domains play a crucial role in the formation of INBs. Since the presence of INBs has been reported for many other hydrophobic/water interfaces, it will be interesting to determine whether these hydrophilic domains also appear at those interfaces. Future investigations should address important issues related to the ordered structures and the interfacial gas hydrate phases, such as details of their molecular configurations, their chemical compositions, their phase diagrams, and their effects on electrochemical reactions and many other processes at hydrophobic/water interfaces. These subjects are fundamentally important and will underlie better understanding of and control over hydrophobic/water interfaces.

**Model for the formation of gas structures at the HOPG/water interface.** Figure 6 depicts a scenario to explain our AFM-based observations. In gas-supersaturated water, dissolved gas molecules form clusters of various sizes (Figure 6a; Supporting Information Additional Discussion 4). Adsorption of clusters larger than a certain size (which remains to be determined) leads to the formation of a circular fluid layer one molecule in thickness at the hydrophobic/water interface (Figure 6b). Adsorption of monomers or small clusters of dissolved gas molecules outside the fluid regions and their bonding arrangement with the interfacial water molecules leads to the gradual nucleation and growth of ordered domains. The fluid layer increases in diameter through further adsorption of gas molecules. The lateral spreading tendency of the fluid layer is hindered when its edge contacts immobile hydrophilic defects, such as substrate step edges or ordered domains. The ordered domains increase in coverage over time and eventually confine the fluid layer within a certain region; the ordered domains may also nucleate at the perimeter of the fluid layer. The subsequent development of the confined fluid region varies with its lateral size. Figure 6c-f illustrate a transformation process in a large confined area. Further adsorption of a certain number of gas molecules onto a large confined fluid regions leads to instability of the 2D layer, which is subsequently transformed into a cap-shaped INB of a smaller diameter through shrinking of the fluid layer (Figure 6d,e). The ordered domains gradually nucleate in the region originally covered by the fluid layer and eventually confine the INB within a small region (Figure 6f).

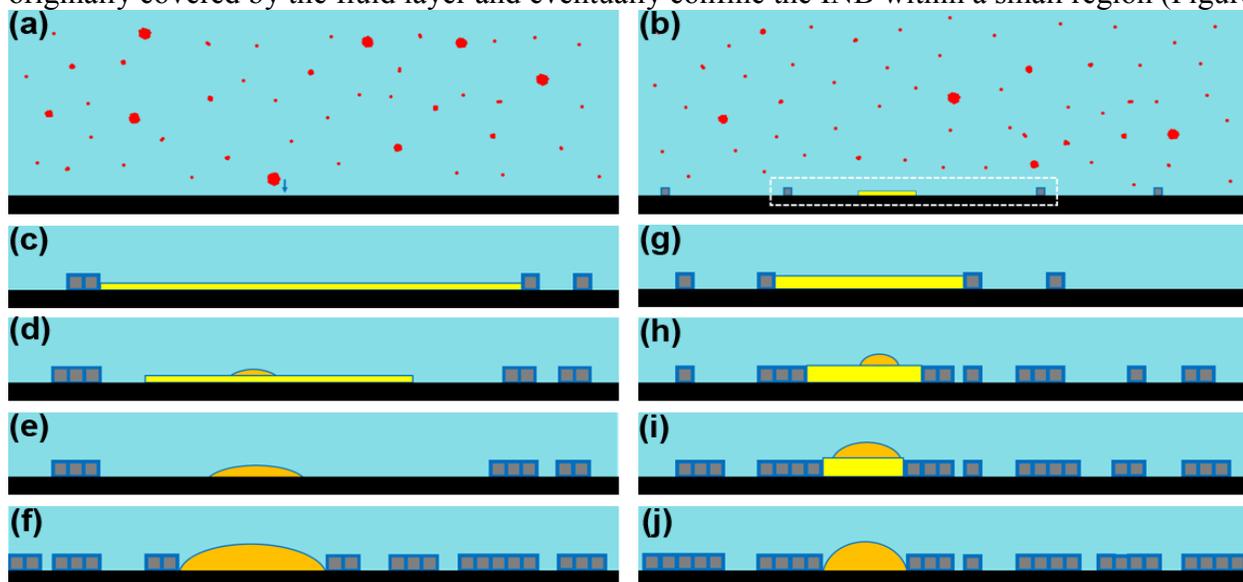

**Figure 6 Scenario for the formation of gas structures at the HOPG/water interface. (a) Dissolved gas molecules (red) are present in clusters of various sizes. (b) Adsorption of a large gas cluster leads to the formation of a circular fluid layer (yellow). Ordered domains (grey) form through the adsorption of monomers and bonding with interfacial water molecules. Subsequent evolution is focused on the solid/water interface (white dashed box). (c)-(f) The case of a large confined fluid region and formation of an INB (orange). (g)-(j) The case of a small confined fluid region and formation of an INB. For clarity, some structures are not represented to scale.**

Figures 6g-j illustrate another scenario in which the diameter of the confined region is small. Incoming gas molecules increase the thickness of the 2D layer (Figure 6g). In addition, the edge of the fluid layer can be gradually converted into ordered domains through rearrangement of the gas molecules at the perimeter of the fluid layers with the surrounding water, gradually decreasing the lateral area of the confined region (Figure 6h,i). At a certain point, a cap-shaped structure appears on the 2D layer (Figure 6h). The cap-shaped structure initially grows in both the lateral and vertical dimensions (Figure 6h,i). When its lateral size reaches the edge of the confined region, further gas adsorption leads to an increase in height only (Figure 6j).

Here, 2D circular layers often appeared near substrate step edges or other line defects of the HOPG substrate (Figures 1 and 2 as well as Figures. S3 and S5), suggesting that the initial fluid layers may be mobile at the interface and tend to persist near these line defects. We expect the mobility of a fluid layer to increase as its lateral size decreases. The irregular region indicated with a blue arrow in Figure 2 may be due to a small fluid layer that moves rapidly around the sparsely scattered ordered domains. A larger fluid layer may be less mobile and more easily confined by scattered ordered domains.

**Nucleation of INBs and gas-supersaturated state.** INBs are generally prepared using gas-supersaturated water[1,6-9]. Interestingly, we also detected the formation of INBs after we deposited water that was rapidly heated to 45 °C (Figure 2). Because the water cools to the temperature of the substrate (room temperature) within 10 s of deposition[40], it becomes undersaturated (60% of the saturation concentration at 25 °C; Methods) soon after deposition. In this investigation, we conducted seven independent experiments with rapidly heated water (one example is shown in Figure 2); INBs were evident in all of them (data not shown). In comparison, deionized water taken directly from water purifiers typically contains oxygen at 60-80% of the saturation concentration at 25 °C. We conducted tens of experiments using fresh deionized water, yet we only observed ordered domains—never the formation of INBs or circular fluid layers (data not shown). Our observations are therefore consistent with previous reports that no INBs are observed at the HOPG/water interface when deionized water is deposited[41,42], that the density of INBs increases with increasing water temperature[40,43,44], and that gas supersaturation is not required for the nucleation of INBs[40].

We were also interested to note that INBs continued to grow in height even in undersaturated water (Figure 2). This continuous increase over several hours to one day also occurred in many other cases, including after the deposition of chilled water (Figures 4 and 5) and after the deposition of rapidly heated water (Figure 2). These observations reflect the strong tendency for dissolved gas molecules to adsorb at hydrophobic/water interfaces, considering that the dissolved gas concentration should decrease over time due to gas adsorption.

Our observations clearly indicate that gas concentration alone does not govern the behaviors of gases dissolved in water and gas accumulation at solid/water interfaces. For each gas concentration, there may be a cluster size distribution that is in thermodynamic equilibrium. A higher gas concentration would have many large clusters. A sudden increase in temperature likely causes the rapid aggregation of dissolved gas molecules into larger clusters (a supersaturated state); the concentration of large clusters increases at the expense of smaller clusters or monomers. Some large clusters are transformed into gas bubbles at the wall of the water container during heating, leading to the removal of a certain amount of gas from water (a reduction in the dissolved gas concentration). However, some large gas clusters remain dissolved in water, and an extended period of time may be required for the gas clusters to relax back to thermodynamic equilibrium after the water cools to room temperature. This scenario explains the formation of INBs using rapidly heated water and the lack of INB formation after the deposition of fresh deionized water, even though the gas concentrations in both cases are similar. Since decompression also leads to the formation of gas bubbles microns in size or larger, similar to the case of a sudden increase in water temperature, we expect that a decrease in pressure may also cause dissolved gas molecules to aggregate into larger clusters. Thus, water may be in a gas-supersaturated state temporarily even though the gas concentration is smaller than the saturation level.

## CONCLUSIONS

This work underlies the importance of microscopic observations of gas-containing structures at solid/water interfaces using highly sensitive AFM. Our unprecedented observations of the details of INB formation reveal an intricate interplay among gas molecules, water, and hydrophobic solids. The observations provide important clues to understand microscopic interactions among gas molecules, water, and hydrophobic solids, which may better understanding of many phenomena at hydrophobic/water interfaces and eventually lead to accurate prediction and control of the interfacial behaviors and properties.

## MATERIALS AND METHODS

**Materials and Sample Preparation** Sample and water preparation. HOPG substrate (12 mm × 12 mm, ZYB grade from Momentive) was freshly cleaved (by peeling off the top layer with Scotch tape) to expose a clean and atomically flat surface. Water was purified with a Milli-Q system (Millipore Corp.) with a resistivity of 18.2 MΩ·cm. Slightly air-supersaturated water was prepared by storing purified Milli-Q water with air in a sealed 50-ml conical centrifuge tube at 4 °C in a refrigerator for several days. Chilled water was either (i) deposited on a HOPG substrate or (ii) heated to 45 °C in a 95 °C hot bath before water deposition (Figure S7). Water was extracted with a pipette and deposited onto the freshly cleaved HOPG substrate, which was kept at room temperature (23-25 °C) in a closed fluid cell equipped with AFM. The oxygen concentration of the chilled water (~4 °C) was measured as 11~12 mg/l using a dissolved oxygen tracer (Lamotte 1761M) (data not shown). When the chilled water was heated to 45 °C, the oxygen concentration was ~5 mg/l (data not shown), or ~60% of the saturation oxygen concentration at 25 °C (8.3 mg/l). Gas bubbles were evident at the inner wall of the centrifuge tube, explaining the decreased oxygen concentration in this rapidly heated water. The water volume in the AFM liquid cell was only 60 µl, and the water temperature was expected to stabilize near room temperature within 10 s after injection into the liquid cell.

**AFM** AFM was performed with a modified Bruker AXS Multimode NanoScope V at room temperature (23-25 °C). FM (Figure S6a) and PF (Figure S6b) modes were used. FM mode was used here because it achieves high force sensitivity in water and more accurately measures the surface profile of INBs and other fluid structures than other operation modes[28,45]; however, more time (10-20 min) is required to tune the imaging conditions before imaging can begin. In addition, stable operation in FM mode may not be achieved for certain tips. With PF mode, AFM can begin sooner (5-10 min) after water deposition and stable imaging can be achieved easily. However, care must be taken when interpreting topographic images acquired in PF mode[28]. We used Si cantilevers (OMCL-AC240TS from Olympus) with a spring constant of 0.7~3.8 N/m, a nominal tip radius of ~10 nm, and a free resonance frequency of ~30 kHz in water. In FM mode, the oscillation amplitude was maintained at 1.0-1.5 nm; the set point was 13-20 Hz.

*Acknowledgment*: We thank support for this work from National Science Council (NSC99-2112-M-001-029-MY3 and NSC102-2112-M-001-024-MY3), the Ministry of Science and Technology (MOST 103-2627-M-001 -011 and MOST 104-2627-M-001-005) of the R.O.C. and Academia Sinica.

# Supporting Information

# Additional Discussion

1. **Driving force for the spreading of the adsorbed gas film at HOPG/water interface.**

Spreading of the adsorbed gas film at the interface is likely driven by gas-graphite interactions, which are much larger than the gas-gas interactions in this system. For example, the energy values of $N_2$–$N_2$, $N_2$–HOPG, $O_2$–$O_2$, and $O_2$–HOPG interactions are approximately 9 meV, 100 meV, 10 meV, and 100 meV, respectively[1], suggesting that the stronger gas-HOPG interactions cause the adsorbed gas film to spread. However, the gas-HOPG interactions are unlikely to be very strong due to the van der Waals nature of the interaction. Thus, an unfavorable interaction (high interfacial tension) between the adsorbed gas structure and water reduces the spreading tendency of the structure, driving it to shrink in the lateral dimension and bulge in the vertical dimension to reduce the interfacial area. The formation of a gas film one molecule in thickness indicates that its interfacial tension with water should be much smaller than the surface tension of water. Interfacial water may form a favorable hydration structure with the smooth gas film. Adsorption of excess gas molecules on the smooth gas film may disrupt the favorable hydration structure and cause the interfacial tension to increase, which becomes the driving force for the transition to a cap-shaped structure at a later time.

2. **Interfacial tension between the fluid structures and water.**

Formation of a complete wetting film at the HOPG/water interface implies that the spreading coefficient[2] $S \equiv \sigma_{HOPG\text{-}water} - (\sigma_{HOPG\text{-}film} + \sigma_{film\text{-}water}) > 0$, where $\sigma_{HOPG\text{-}water}$, $\sigma_{HOPG\text{-}film}$, and $\sigma_{film\text{-}water}$ represent the interfacial tension of HOPG/water, gas-film/HOPG, and gas-film/water, respectively. As estimated below, $\sigma_{HOPG\text{-}water}$ is smaller than the surface tension of water ($\sigma_{air\text{-}water}$ = 72 mN/m). Although the value of $\sigma_{HOPG\text{-}film}$ is not known, it should be small due to the lack of strong bonding between gas molecules (nitrogen or oxygen) and the HOPG substrate. Thus $\sigma_{film\text{-}water} < \sigma_{HOPG\text{-}water} < \sigma_{air\text{-}water}$.

According to Young's law, $cos\theta_{eq} = (\sigma_{\alpha\gamma} - \sigma_{\beta\gamma})/\sigma_{\alpha\beta}$ (refer to the figure below).

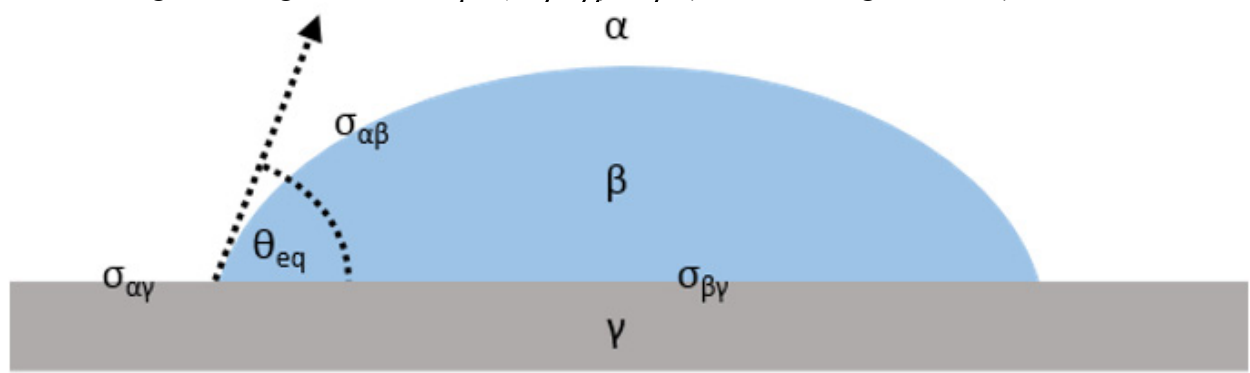

Since the contact angle of water on freshly cleaved HOPG is 65°[3] we estimate $\sigma_{HOPG\text{-}water}$ based on Young's law and the surface tensions of water and HOPG. The surface tension of water, $\sigma_{air\text{-}water}$, is 72 mN/m, and that of HOPG is 35-55 mN/m[4,5]. Thus, $\sigma_{HOPG\text{-}water}$ is estimated to be 25 mN/m or smaller.

For an interfacial nanobubble (INB) at a solid/water interface, the typical contact angle is 5°-25°. The interfacial tension between an INB and water, $\sigma_{INB\text{-}water}$, should be somewhat larger than $\sigma_{film\text{-}water}$, but still considerably smaller than the surface tension of water. The small interfacial tension would make the INB highly deformable and tend to adopt a rather flat morphology, which is very different from the stiff semispherical shape typically seen for a gas bubble of micron size or larger at a solid/water interface. Small interfacial tension also contributes to lower free energies of INBs and explain the high stability of INBs.

**3. Nanobubbles as the micronuclei for gas bubble formation in water**

It has been theoretically predicted that a vapor cavity will form only when the liquid is under extremely large tension[6]. However, these large tensile strengths have never been observed, leading to the introduction of the idea of micronuclei. Plesset stated in 1969[7] that small gas bubbles cannot be the persistent nuclei for the formation of micro- or macro-bubbles in water due to their short lifetimes[6]. Plesset also conjectured[7] that dissolved gases may form adsorbed films of gas on hydrophobic surfaces; these films would have a low surface tension and could act as nucleation centers for bubble formation in water. Arieli and Marmur[8] proposed that surface nanobubbles may be the gas micronuclei responsible for the bubbles that cause decompression sickness, but the mechanism was not known. Our study indicates that INBs are gas aggregates (or gas condensates) with a low interfacial tension with water, providing insight into the nucleation of gas bubbles in water.

**4. Gas cluster distribution in water**

The two types of gas-containing interfacial structures observed in the present investigation, ordered domains and fluid structures, may originate from different gas configurations in bulk water. During numerous atomic force microscopy (AFM) studies of the HOPG/water interface that we conducted under ambient conditions[9-11], we typically observed INBs or fluid layers when air-saturated or gas-supersaturated water was deposited. Ordered domains appeared in saturated or super-saturated water as well as in partially degassed water[9]. In partially degassed water, it is expected that dissolved gas molecules are present mainly in the form of well-dispersed monomers; thus, ordered domains form through self-assembly of individual monomers or small clusters at the interface. In air-supersaturated water, a subset of the dissolved gas molecules probably aggregates into large clusters. Theoretical studies have indicated that small non-polar molecules can aggregate into clusters nanometres in size in bulk water[12,13]. Nitrogen and oxygen, the two major components of air, are non-polar, suggesting that they may aggregate into large clusters with a diameter >1 nm when the gas concentration is near or above the saturation level. Adsorption of a large cluster at the HOPG/water interface may lead to the formation of a circular fluid layer that later evolves into a cap-shaped INB.

# Supporting Figures

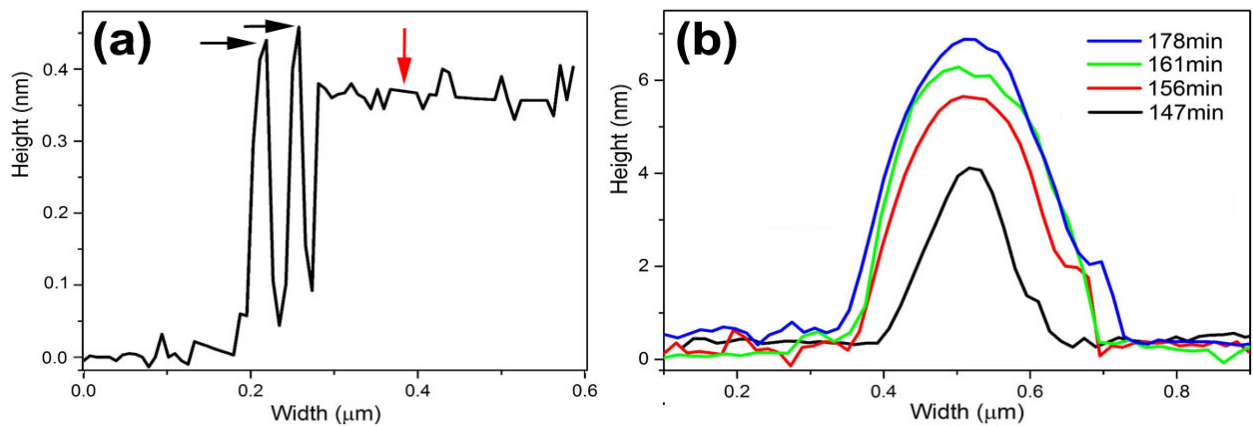

Figure S1 Height profiles measured in Figure 1 of the main text. (a) Height profiles along the dashed line in Figure 1b of the main text. Black arrows indicate the profile across ordered domains; the red arrow indicates the profile across a fluid layer. (b) Evolution of the height profiles of the INB in region 4 at different times.

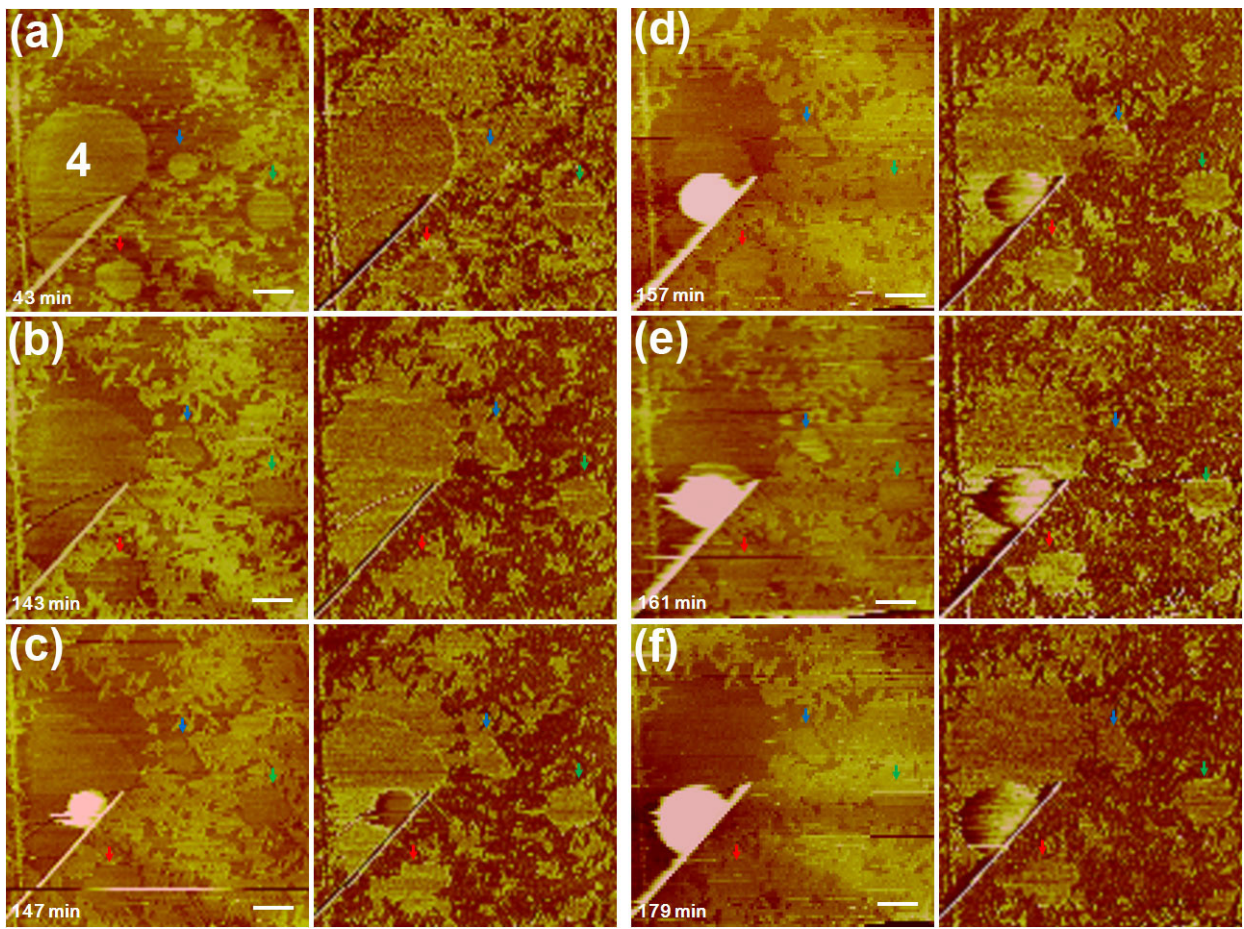

Figure S2  AFM images of the transformation process around region 4 in Figure 1 of the main text. The left and right columns of each panel depict the topographic and dissipation images, respectively. Three small fluid layers are indicated with colored arrows. The fluid layers exhibited a lower height than the ordered domains in (a-c), but they grew in height at later stages and could not be distinguished easily from the surrounding ordered domains in topographic images. The ordered domains displayed a lower dissipation than the fluid layers and bare HOPG regions; thus, the dissipation images enabled us to better distinguish the ordered domains from other regions. Scale bar, 200 nm.

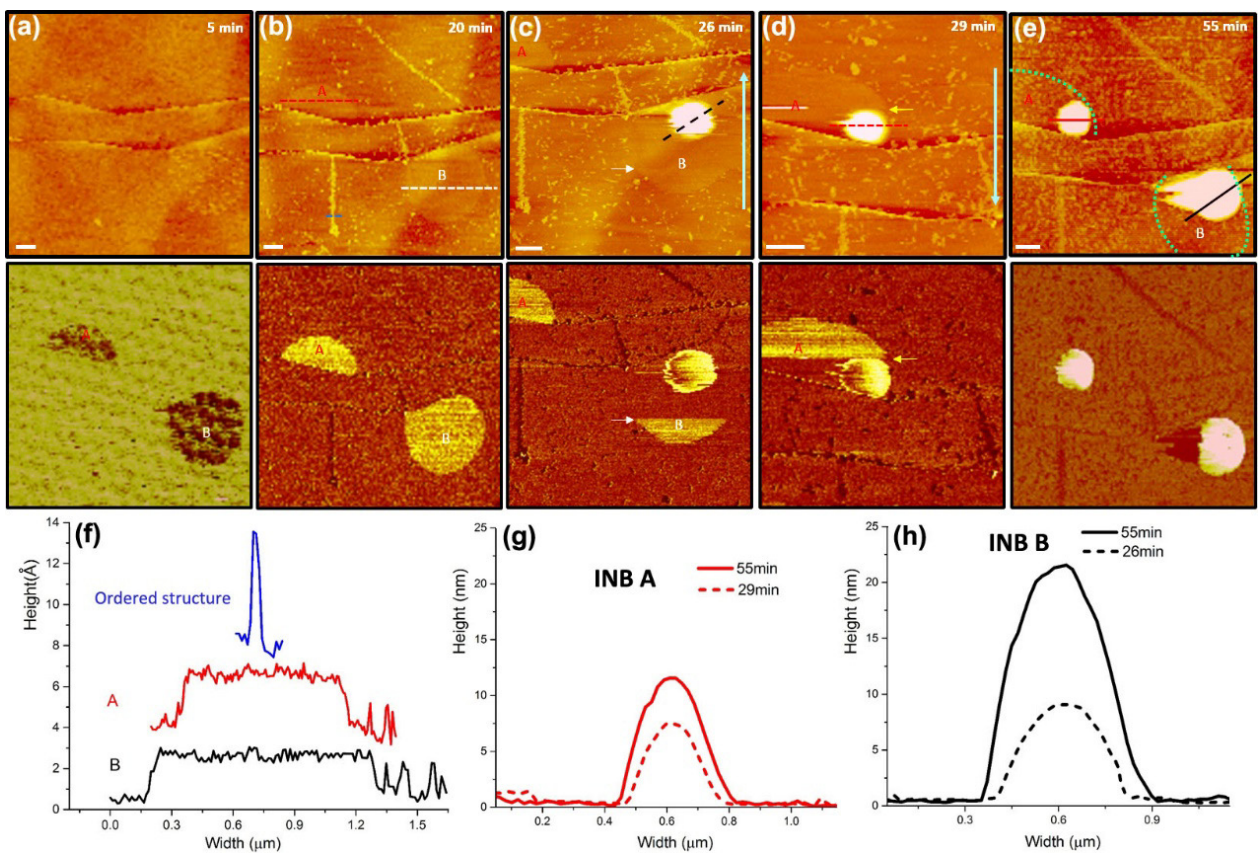

Figure S3 AFM of the transition of large fluid layers to INBs at a HOPG/water interface. (a) Topographic (top) and adhesion (bottom) images acquired in PF mode at $t$=5 min. Scale bar in all panels, 250 nm. The adhesion map provides considerably higher contrast for the two circular layers (indicated with A and B) than does the topographic map. Adsorption of circular layers had already occurred at $t$=5 min. Imaging was subsequently switched to FM mode; (b)-(e) contain the topographic (top) and dissipation (bottom) images acquired with this mode at various times. The topographic images also show two circular layers ~0.3 nm in thickness, and the dissipation maps exhibit strong contrast for the fluid regions relative to other parts of the interface. The small bright speckles in the topographic images are ordered domains; they appear darker than the bare HOPG surface in the dissipation images. In (c), circular layer B disappears suddenly at the scan line indicated by a white arrow, and a cap-shaped nanostructure appears at subsequent scan lines. Note that the images were acquired through raster scanning; blue arrows at the side of the topographic images indicate the slow scan direction. Each frame took ~3 min to acquire. The time to scan one line was 1 s. In (d), circular layer A suddenly transforms into a cap-shaped nanostructure. In (e), the two cap-shaped INBs grow larger in both the vertical and lateral dimensions. In addition, small ordered domains (bright speckles) appear inside the regions that were originally covered by the thin layers (outlined by green dashed lines). (f) Height profiles in (b). (g) Height profile across INB A at two time points. (h) Height profile across INB B at two time points. Both INBs exhibited growth in the vertical and lateral directions.

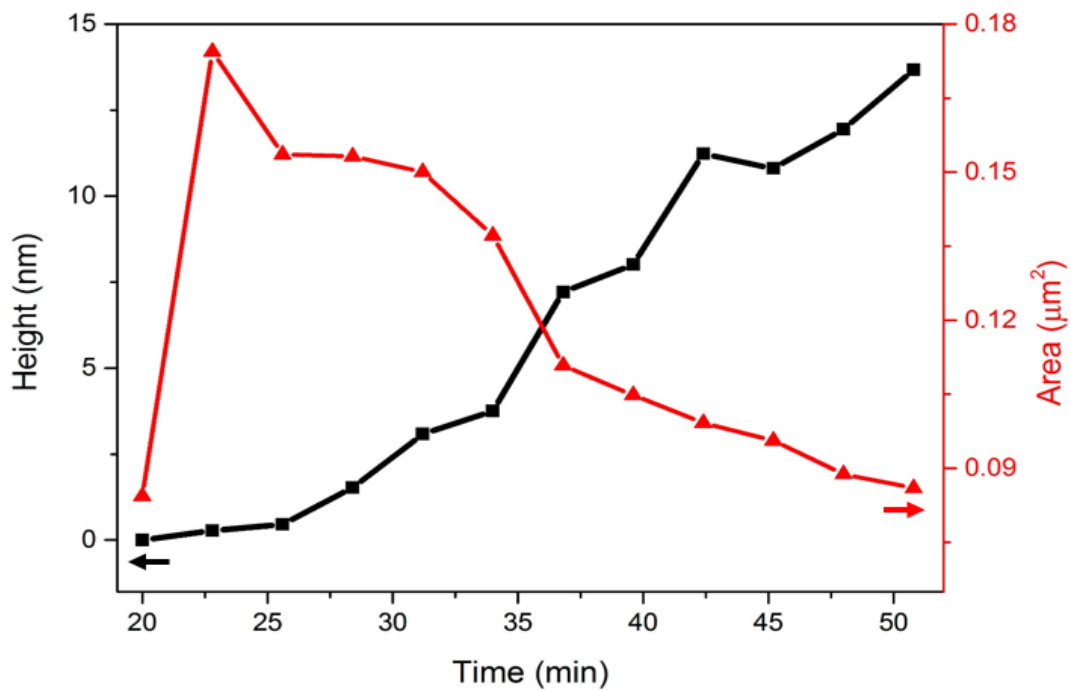

Figure S4 Evolution of lateral area and apparent height over time for the fluid structure indicated with a white arrow in Figure 2 of the main text. Lateral area was measured from the stiffness maps and apparent height was measured from the highest point of the height profiles.

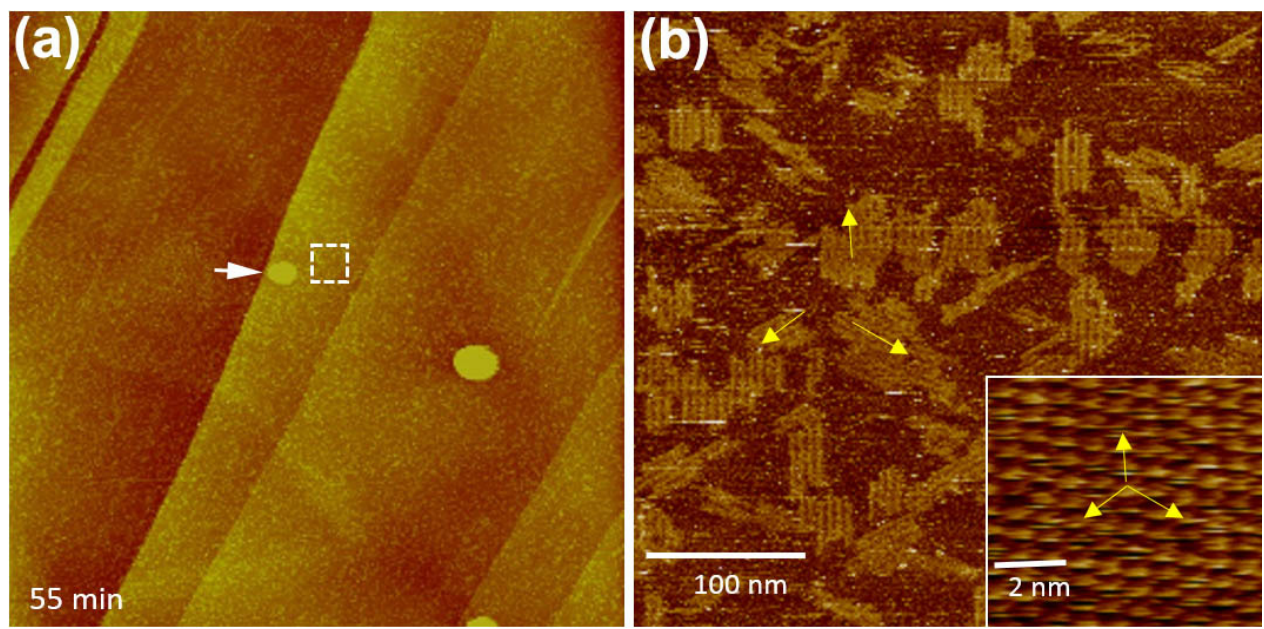

Figure S5  Topographic (PF, 250 pN) images of structures at the HOPG-water interface acquired at $t=$ 55 min (a) and $t=$ 71 min (b). The INB indicated with a white arrow serves as a marker for comparison with images shown in Figure 2 of the main text. Two additional INBs were observed outside the scan area shown in Figure 2 of the main text. The image shown in (b) was acquired in the outlined region in (a). The yellow arrows in (b) indicate the three orientations of the row-like structures, which are parallel with the lattice orientations of the HOPG substrate, zig-zag directions. The HOPG lattice was acquired at a high loading force with the contact mode (inset).

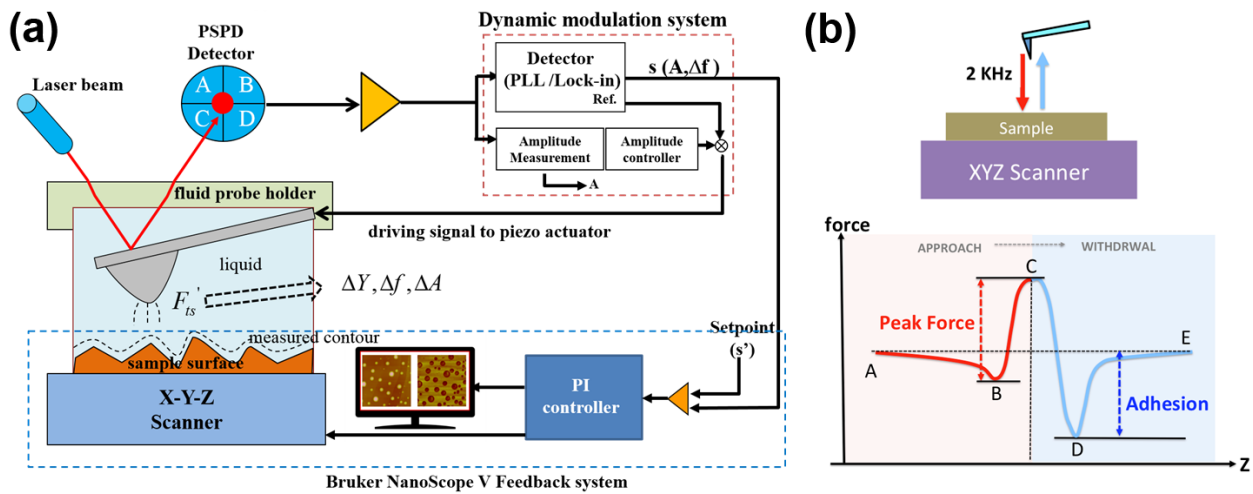

Figure S6 AFM operation. (a) Schematic of the Multimode NanoScope V modified for FM mode. Oscillation of the cantilever was driven using a dynamic-modulation system comprised of a lock-in unit/phase-lock-loop (PLL) unit (Nanonis OC4 Station from SPECS) and a signal-access module (Bruker AXS). The PLL unit was employed to track the resonance frequency of the vibrating cantilever. The resonance frequency shift ($\Delta f$) was used as the feedback input signal of a proportional-integral controller to obtain topographic images. The driving amplitude, which we controlled with the PLL in order to maintain a constant amplitude of cantilever oscillation during scanning, was measured and recorded simultaneously. The amplitude value was typically considered to be the energy dissipation signal[14]. A dissipation map was acquired along with the topographic images. (b) Illustrations of PF mode. Vertical piezo movement results in cycles of approaching and retracting traces in which the tip makes intermittent contact with the sample surface, yielding force-distance curves. Topography information is obtained from the height correction performed by the feedback loop to keep a constant "peak" of force, while the slope of the contact region determines the stiffness of the sample at each pixel. Adhesion and deformation can also be extracted from the force curves. Our previous studies indicated that snap-in usually occurs when the AFM tip touches a fluid region such as an INB. The tip must penetrate the structures to a certain depth to offset the attractive snap-in force to reach a positive preset peak force[12]; thus, the penetration depth, which varied with the hydrophobicity of the AFM tip[16], should be added to the height profile of the fluid region obtained from the topography images obtained in PF mode. If a fluid structure is thinner than the penetration depth, it cannot be seen in the topographic images because the tip traces the profile of the stiff structure underneath. Snap-in, which is caused by the hydrophobicity of the tip apex and the fluid nature of the structure under study[15], does not strongly affect topographic images acquired in FM mode[1], probably because the feedback signal (the resonance frequency shift) is related to the force gradient rather than to the interaction force itself. A sharp increase in the resonance frequency often occurs when the tip touches a fluid structure[1].

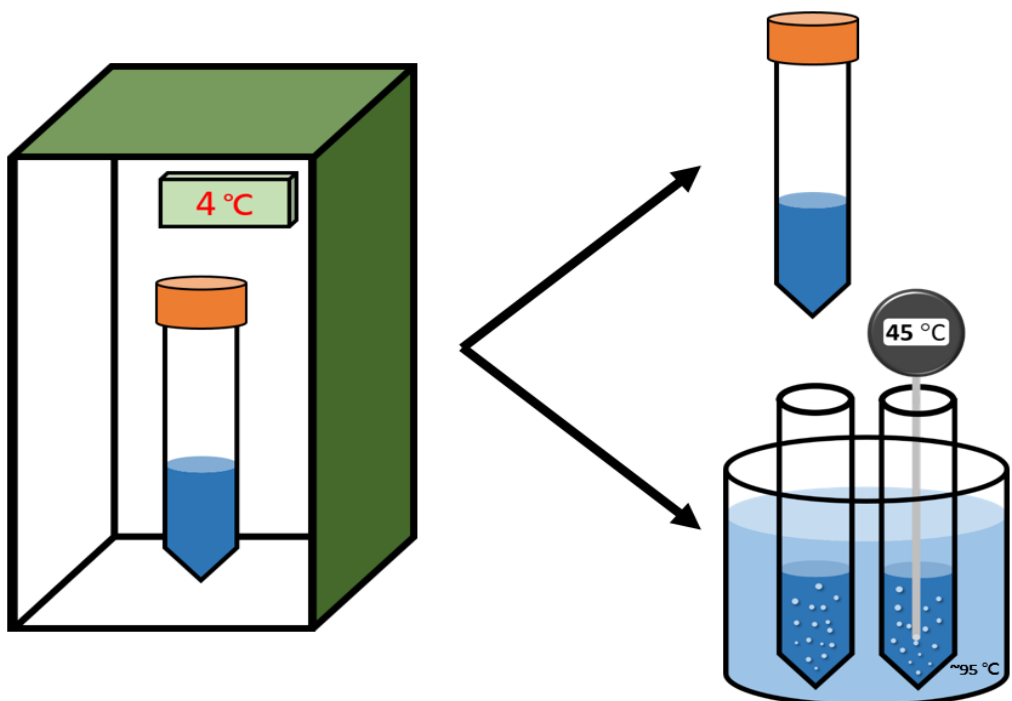

Figure S7 Water preparation before AFM. Purified deionized water was first stored with air in a sealed centrifuge tube at 4°C in a refrigerator for several days. The chilled water could be either (i) deposited on a HOPG substrate in a AFM liquid cell at room temperature or (ii) heated to 45°C in a 95°C hot water bath before deposition on a HOPG substrate in a liquid cell.

# Supplementary References